\documentclass[prl,aps,epsfig,twocolumn]{revtex4}
\usepackage{amssymb}

\begin{document}

%\textbf{Fermi Surface Splitting in Na$_x$CoO$_2$} 
\textbf{ Comment on \textquotedblleft Low-Lying Quasiparticle States and Hidden
Collective Charge Instabilities in Parent Cobaltate Superconductors"}

Qian \textit{et al}\cite{1} recently reported angular-resolved photoemission spectroscopy (ARPES) measurements for
Na$ _{0.8}$CoO$_{2}$ that show two concentric Fermi surfaces (FS) split by a $\delta k_{F}$ that varies
by a factor of three around the Brillouin zone (BZ). The surfaces occupy 70$\pm 5$\% of the full 2D BZ and were
interpreted as the bonding and antibonding splitting (BAS) of the $ a_{1g}$ bands, with an unspecified effect of
magnetic ordering. Below we show that this interpretation is not possible, and, in fact, no valid intepretation of
the observed spectra in terms of the \textit{bulk} electronic structure of Na$_{0.8}$CoO$_{2}$ can be found.

The two formulas per unit cell of Na$_{x}$CoO$_{2}$ double all
bands in the Brillouin zone, including the observed $a_{1g}$ band, formed by
Co $d_{z^{2}-1}$ orbitals. Symmetry mandates that the BAS is zero at $
k_{z}=\pi /c,$ but does not, in general, prescribe its $k_{z}$ and $k_{x,y}$
dependence. The crystal structure, however, only allows for sizeable hopping
between Co planes via connecting O-O dumbells. This fact and the $
d_{z^{2}-1}$ symmetry of the orbitals give rise to two corollaries: (a) the
BAS is proportional to $\cos k_{z}c,$ \textit{i.e., } maximal at $k_{z}=0$
and (b) the $k_{x,y}$ dependence of the BAS is, to very good accuracy,
proportional to $t_{\mathrm{O-O}}t_{\mathrm{Co-O}}^{2}\sum_{i}\cos \mathbf{A}
_{i}\mathbf{\cdot k}_{xy},$ where $\mathbf{k}_{xy}$ is the in-plane vector
and \textbf{A}$_{1,2,3}$\textbf{\ } are the three nearest-neighbor Co-O
vectors. Note that this functional form is not related to LDA or any other
approximation (the value of the prefactor is), but only to the symmetry of
the underlying Hamiltonian. At the edge of the first BZ this expression
provides a maximum BAS angular anisotropy of $\delta k_{F}/\langle \delta
k_{F}\rangle <$15\%\thinspace\ while at the $k_{F}$ measured in Ref.
\onlinecite{1} $\delta k_{F}/\langle \delta k_{F}\rangle <$ 2\%,\thinspace\
to be compared with an observed factor of three.
The discrepancy of three orders of magnitude leaves no doubt that
the observed splitting is not the bulk BAS.

Contrary to a claim in Ref. \onlinecite{1}, it is not the AFM ordering that
\textquotedblleft leads to canonical doubling of the unit
cell\textquotedblright ; it is doubled already without magnetism and
the AFM only enhances the existing BAS. The total splitting is $\sqrt{\tau
^{2}+\Delta _{ex}^{2}},$ where $\tau $ is the nonmagnetic BAS and $\Delta
_{ex}$ is the exchange splitting, which can, in principle, depend on $
\mathbf{k}_{xy}.$ However, this interpretation can also be safely excluded:
doubly degenerate AFM FSs would contain $0.7$ holes/formula, not
0.2, as required by Na content. Qian $et$ $al$ argue that \textquotedblleft
the 2D Luttinger count is not applicable to the FS of highly doped
cobaltates\textquotedblright . However, for any practical purpose, it is:
the cosine dependence of the BAS mandates that without AFM the Luttinger
theorem (LT) is satisfied at each $k_{z}$ \textit{separately}. With AFM it
is satisfied within $\tau ^{2}/\Delta _{ex}^{2},$ and to explain the
large splitting anisotropy one has to assume that $\tau \ll \Delta _{ex}$.

Thus, the FS observed in Ref. \onlinecite{1} cannot represent the bulk FS. We now speculate on what kind of surface
effects could help explain this observation. Since a polar surface cannot be stable\cite{George}, the termination
layer in Na$_{x}$CoO$_{2}$ cannot be Na$_{x},$ as in the bulk. Let it be Na$_{y}$ (0$\leq y\leq x)$. Conditions of
nonpolarity and total neutrality imply that the top CoO$_{2}$ layer then carries a charge of $z=-(x/2+y),$ or
$1-x/2+y$ holes.  The implicit assumption in Ref. \onlinecite{1} that the outermost CoO$_2$ layer
has the bulk hole concentration,  CoO$_{2}^{-x}$, would require the surface Na concentration to be $y=x/2$.
As we have argued, this assumption leads to a severe violation of the LT. Moreover, 
since the termination layer is now Na$_{x/2}$, the first and second CoO$_{2}$ layers see different Na
potentials. If ARPES were probing the top two CoO$_{2}$ layers, two spin-split bands would be observed for each
layer, totalling four bands for $x=0.8,$ (there is no known mechanism that would selectively suppress one spin, but
not the other in non-spin-polarized PES), and two bands for $x<0.6$ (even in the absence of BAS and AFM). That
neither is the case proves that only one CoO$_{2}$ layer is probed.

For $x=0.8,$ the ARPES hole count is incompatible with Na$_{x/2}$ termination.
It is, however, approximately compatible with $y=0$ termination (no Na on the
surface) where the top layer is CoO$_{2}^{-0.4}$ (0.6
holes), roughly agreeing with 0.70$\pm 0.05$. The observed
splitting may be ascribed to exchange, allowing at least for some possibility of explaining the angular
anisotropy. Note that although the measured  magnetic moment of 0.13$\pm 0.02
$ (at $x=0.82$, Ref. \onlinecite{Sybel}) implies a larger FS splitting, the
surface layer may be less polarized than the bulk.
On the other hand, the \textit{bulk }LT is fulfilled for the data in Fig. 2d
of Ref. \onlinecite{1} for $x\lesssim 0.6,$ compatible (assuming only the top
layer is probed), with Na$_{x/2}$ termination and a CoO$_{2}^{-x}$
surface layer. While the electronic structure of this layer will not
be identical to the bulk, its doping level is.

To summarize, we have shown that the observed\cite{1} Fermi surfaces cannot
represent the bulk electronic structure due to severe restrictions on the
bonding-antibonding splitting anisotropy imposed by the crystal symmetry,
and the impossibility of satisfying the LT, either with
or without the AFM spin density wave. This conclusion is not model-specific
and follows from general symmetry considerations. We also point out the
impossibility of creating a nonpolar surface while maintaining the bulk Na
concentration on the surface. The ARPES data of Ref. \onlinecite{1} appear to be
fully understandable under the assumption that only the top CoO$_{2}$ layer
is probed, with a magnetically ordered surface with \textit{
no} Na termination for bulk doping $x\gtrsim 0.6,$ and a nonmagnetic surface
with \textit{half} Na termination, Na$_{x/2},$ for  $x\lesssim 0.6$\cite
{Alloul}. The two observed FSs at $x=0.8$ then correspond to the two spin
directions. We emphasize however, that these are only possible explanations and that the 
main purpose of our 
Comment is to show what {\it cannot} be rather than to speculate about what {\it can} be.

I.I. Mazin$^{1}$, M.D. Johannes$^{1}$, and G.A. Sawatzky$^{2}$ \newline
$^{1}${\footnotesize Naval Research Laboratory, Washington, DC 20375, U.S.A,
and }$^{2}${\footnotesize University of British Columbia, Vancouver, Canada}

\end{document}